\documentclass{bmcart}

\usepackage{xr-hyper} % reference SI file
\usepackage[utf8]{inputenc} %unicode support
\usepackage{graphicx}
\usepackage{booktabs}
\usepackage{amsmath, amssymb, amstext, amsthm, amsfonts, bm, float}
\usepackage{bbm}
\usepackage{subcaption}
\usepackage{enumitem}
\usepackage{hyperref}

\makeatletter
\newcommand*{\addFileDependency}[1]{% argument=file name and extension
  \typeout{(#1)}
  \@addtofilelist{#1}
  \IfFileExists{#1}{}{\typeout{No file #1.}}
}
\makeatother

\newcommand*{\myexternaldocument}[1]{%
    \externaldocument{#1}%
    \addFileDependency{#1.tex}%
    \addFileDependency{#1.aux}%
}
\myexternaldocument{./supplementary_material}

\usepackage{color}
\definecolor{redcolor}{rgb}{.7,0.,0.}
\DeclareMathOperator*{\argmax}{arg\,max}

\startlocaldefs
\endlocaldefs

\begin{document}

%%% Start of article front matter
\begin{frontmatter}

\begin{fmbox}
\dochead{Research}


\title{Multilayer Networks for Text Analysis with Multiple Data Types}

%%%%%%%%%%%%%%%%%%%%%%%%%%%%%%%%%%%%%%%%%%%%%%
%%                                          %%
%% Enter the authors here                   %%
%%                                          %%
%% Specify information, if available,       %%
%% in the form:                             %%
%%   <key>={<id1>,<id2>}                    %%
%%   <key>=                                 %%
%% Comment or delete the keys which are     %%
%% not used. Repeat \author command as much %%
%% as required.                             %%
%%                                          %%
%%%%%%%%%%%%%%%%%%%%%%%%%%%%%%%%%%%%%%%%%%%%%%

\author[
   addressref={aff1},                   % id's of addresses, e.g. {aff1,aff2}
   corref={aff1},                       % id of corresponding address, if any
   %noteref={n1},                        % id's of article notes, if any
   email={eduardo.altmann@sydney.edu.au christopherhyland95@gmail.com}   % email address
]{\inits{CCH}\fnm{Charles C.} \snm{Hyland}}
\author[
   addressref={aff1},
   email={???}
]{\inits{YT}\fnm{Yuanming} \snm{Tao}}
\author[
   addressref={aff1},
   email={lamiae.azizi@sydney.edu.au}
]{\inits{LA}\fnm{Lamiae} \snm{Azizi}}
\author[
   addressref={aff2},
   email={mgerlach@wikimedia.org}
]{\inits{MG}\fnm{Martin} \snm{Gerlach}}
\author[
   addressref={aff3,aff4},
   email={peixotot@ceu.edu}
]{\inits{TPP}\fnm{Tiago P.} \snm{Peixoto}}
\author[
   addressref={aff1},
   email={eduardo.altmann@sydney.edu.au}
]{\inits{EGA}\fnm{Eduardo G.} \snm{Altmann}}
%%%%%%%%%%%%%%%%%%%%%%%%%%%%%%%%%%%%%%%%%%%%%%
%%                                          %%
%% Enter the authors' addresses here        %%
%%                                          %%
%% Repeat \address commands as much as      %%
%% required.                                %%
%%                                          %%
%%%%%%%%%%%%%%%%%%%%%%%%%%%%%%%%%%%%%%%%%%%%%%

\address[id=aff1]{%                           % unique id
  \orgname{School of Mathematics and Statistics, The University of Sydney}, % university, etc
  \street{NSW},                     %
  \postcode{2006}                                % post or zip code
  \city{Sydney},                              % city
  \cny{Australia}                                    % country
}
\address[id=aff2]{%
  \orgname{Wikimedia Foundation}
  %\street{D\"{u}sternbrooker Weg 20},
  %\postcode{24105}
  %\city{Kiel},
%   \cny{United States}
}
\address[id=aff3]{%
  \orgname{Department of Network and Data Science, Central European University},
  \street{Quellenstraße 51},
  \postcode{1100}
  \city{Vienna},
  \cny{Austria}
}
\address[id=aff4]{%
  \orgname{Department of Mathematical Sciences, University of Bath},
  \street{Claverton Down},
  \postcode{BA2 7AY}
  \city{Bath},
  \cny{United Kingdom}
}

%%%%%%%%%%%%%%%%%%%%%%%%%%%%%%%%%%%%%%%%%%%%%%
%%                                          %%
%% Enter short notes here                   %%
%%                                          %%
%% Short notes will be after addresses      %%
%% on first page.                           %%
%%                                          %%
%%%%%%%%%%%%%%%%%%%%%%%%%%%%%%%%%%%%%%%%%%%%%%

% \begin{artnotes}
% %\note{Sample of title note}     % note to the article
% \note[id=n1]{Equal contributor} % note, connected to author
% \end{artnotes}

\end{fmbox}% comment this for two column layout

%%%%%%%%%%%%%%%%%%%%%%%%%%%%%%%%%%%%%%%%%%%%%%
%%                                          %%
%% The Abstract begins here                 %%
%%                                          %%
%% Please refer to the Instructions for     %%
%% authors on http://www.biomedcentral.com  %%
%% and include the section headings         %%
%% accordingly for your article type.       %%
%%                                          %%
%%%%%%%%%%%%%%%%%%%%%%%%%%%%%%%%%%%%%%%%%%%%%%

\begin{abstractbox}

\begin{abstract} % abstract
We are interested in the widespread problem of clustering documents and finding topics in large collections of written documents in the presence of metadata and hyperlinks.  To tackle the challenge of accounting for these different types of datasets, 
we propose a novel framework based on Multilayer Networks and Stochastic Block Models. 
The main innovation of our approach over other techniques is that it applies the same non-parametric probabilistic framework to the different sources of datasets simultaneously. The key difference to other multilayer complex networks is the strong unbalance between the layers, with the average degree of different node types scaling differently with system size. We show that the latter observation is due to generic properties of text, such as Heaps' law, and strongly affects the inference of communities.  We present and discuss the performance of our method in different datasets (hundreds of Wikipedia documents, thousands of scientific papers, and thousands of E-mails) showing that taking into account multiple types of information provides a more nuanced view on topic- and document-clusters and increases the ability to predict missing links.

% \parttitle{First part title} %if any
% Text for this section.

% \parttitle{Second part title} %if any
% Text for this section.
\end{abstract}

%%%%%%%%%%%%%%%%%%%%%%%%%%%%%%%%%%%%%%%%%%%%%%
%%                                          %%
%% The keywords begin here                  %%
%%                                          %%
%% Put each keyword in separate \kwd{}.     %%
%%                                          %%
%%%%%%%%%%%%%%%%%%%%%%%%%%%%%%%%%%%%%%%%%%%%%%

\begin{keyword}
\kwd{Stochastic Block Models}
\kwd{Multilayer Networks}
\kwd{Natural Language Processing}
\kwd{Complex Systems}
\kwd{Data Science.}
\end{keyword}

% MSC classifications codes, if any
%\begin{keyword}[class=AMS]
%\kwd[Primary ]{}
%\kwd{}
%\kwd[; secondary ]{}
%\end{keyword}

\end{abstractbox}
%
%\end{fmbox}% uncomment this for twcolumn layout

\end{frontmatter}

%%%%%%%%%%%%%%%%%%%%%%%%%%%%%%%%%%%%%%%%%%%%%%
%%                                          %%
%% The Main Body begins here                %%
%%                                          %%
%% Please refer to the instructions for     %%
%% authors on:                              %%
%% http://www.biomedcentral.com/info/authors%%
%% and include the section headings         %%
%% accordingly for your article type.       %%
%%                                          %%
%% See the Results and Discussion section   %%
%% for details on how to create sub-sections%%
%%                                          %%
%% use \cite{...} to cite references        %%
%%  \cite{koon} and                         %%
%%  \cite{oreg,khar,zvai,xjon,schn,pond}    %%
%%  \nocite{smith,marg,hunn,advi,koha,mouse}%%
%%                                          %%
%%%%%%%%%%%%%%%%%%%%%%%%%%%%%%%%%%%%%%%%%%%%%%

%%%%%%%%%%%%%%%%%%%%%%%%% start of article main body
% <put your article body there>

%%%%%%%%%%%%%%%%
%% Background %%https://fr.overleaf.com/project/600d8f8e2750e7e3ac87c3cd
%%
\section{Introduction}
A widespread problem in modern Data Science is how to combine multiple data types such as images, text, and numbers in a meaningful framework \cite{kedem2017, Costanedo2013,Zhu2013,Kivela2013,Zanin2016}. The traditional approach to tackle this challenge is to construct machine learning pipelines in which each data type is treated separately -- sequentially or in parallel -- and the partial results are combined at the end of the procedure \cite{Breck2019, Oleary2020}. 
There are two problems with such a procedure. First, it leads to the development of ad-hoc solutions that are highly contingent on the dataset in question \cite{Arun2010, Cao2009}. Second, each model is trained independently from one another, meaning that the relationships between the different types of data are not taken into account~\cite{Valles-Catala2016,Peixoto2015Mesoscale}. These problems show the need of developing a unified statistical framework applied simultaneously to the different types of data~\cite{Peixoto2019bayesian}.

\begin{figure}[h!]
  \includegraphics[width=7.5cm]{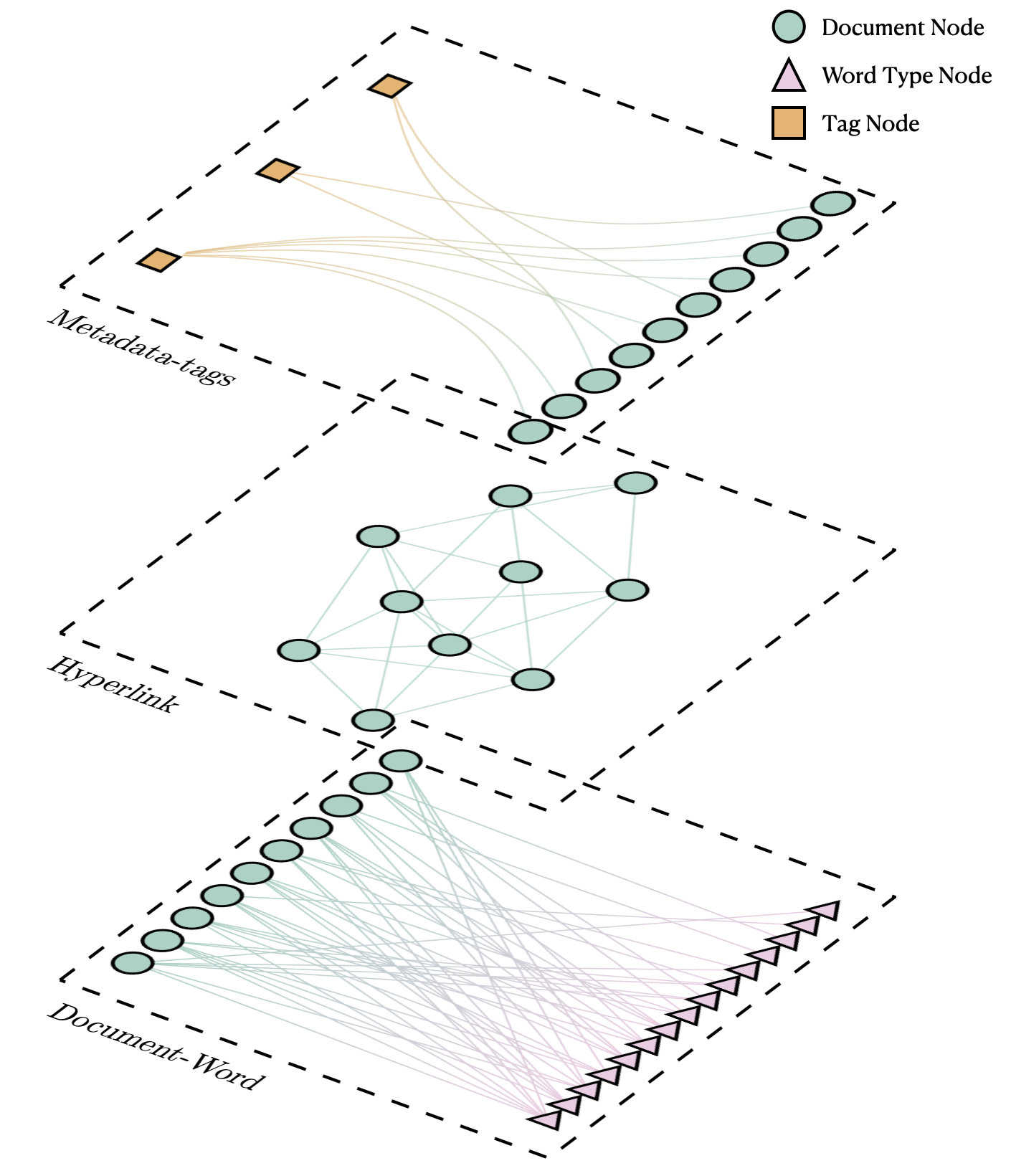} 
  \caption{\textbf{Different views on the relationship between written documents.} Lower layer: a bipartite multi-graph of documents (circles) and word types (triangles), links correspond to word tokens. Middle layer: directed graph of documents (e.g., hyperlinks in Wikipedia or citations in Scientific Papers). Upper layer: a bipartite graph of documents and tags (squares), used to classify the documents.}\label{fig:multilayer-sbm}
\end{figure}

 In this paper, we investigate the problem of clustering and finding topics in collections of written documents for which additional information is available as metadata and as hyperlinks between documents. We obtain a unified statistical framework to this problem by mapping it to the problem of inferring groups in multilayer networks. The key design for the unified framework proposed here is inspired by the connections~\cite{Zhu2013,Ball2011, Lancichinetti2015, Gerlach2018} between the problems of identifying (i) topics in a collection of written documents (i.e. topic modeling)~\cite{Blei2012} and (ii) communities in complex networks (i.e. community detection)~\cite{Fortunato2010}. In particular, Ref.~\cite{Gerlach2018} shows that both problems can be tackled using Stochastic Block Models (SBM) \cite{Zhu2013,Peixoto2019bayesian,Bouveyron2016,Holland1983, Karrer2011,  Hastings2006,larremore} and that SBMs, previously applied to find communities in complex networks, outperform and overcome many of the difficulties of the most popular unsupervised methods to infer structures from large collections of texts (topic modelling methods such as the Latent Dirichlet Allocation~\cite{Blei2003} and its generalizations). However, these approaches have been applied only to the textual part of collections of documents, ignoring additional information available about them. For instance, in datasets of scientific publications, one would consider only the text of the articles but not the citation networks (used in traditional community detection methods \cite{Fortunato2010}) or other metadata (such as the journal or bibliographical classification) ~\cite{Hric2016, Newman2015}. We propose here an extension of Ref.~\cite{Gerlach2018} and show how the diversity of information typically available about documents can be incorporated in the same framework by using multilayer SBMs \cite{Zhu2013, Valles-Catala2016,Peixoto2015Mesoscale}. As illustrated in Fig.~\ref{fig:multilayer-sbm}, in addition to the bipartite Document-Word layer discussed in Ref.~\cite{Gerlach2018}, here we incorporate a Hyperlink layer connecting the different written documents and a Metadata-Document layer that incorporates tags and other metadata. The key difference to other multilayer networks~\cite{Kivela2013}, as explored in Sec.~\ref{sec.2} below, is that statistical laws~\cite{Altmann2015} governing the frequency of words on documents leave fingerprints on the density of the different network layers. Our investigations in different datasets, reported in Sec.~\ref{sec.3}  for collection of Wikipedia articles and in the Supplementary Information for three other datasets, reveal that the proposed multilayer approach leads to improved results when compared to both the topic modelling approach of \cite{Gerlach2018} and the usual community detection of (hyperlink) networks. Our approach leads to a more nuanced view on the communities of documents, generates a list of topics associated to the communities, and improves the link-prediction capabilities when compared to the hyperlink network alone \cite{Guimera2009}. The details on our methods can be found in the appendices, Supplementary Information, and in the repository~\cite{Graphtool}.

\section{Multiple data sources as multilayer networks}\label{sec.2}

In this section we introduce the general methodology of our paper:
we introduce the types of data we are interested in (Sec. ~\ref{sec.setting}), we show how they can be represented as a multilayer network and discuss the properties of these networks (Sec.~\ref{sec.data}), and we describe how they can be modelled using Stochastic Block Models (Sec.~\ref{sec.sbm}). 

\subsection{Setting: Multiple Data Sources}\label{sec.setting}

We consider a collection of $d=1, \ldots,D$ documents and we are interested in clustering and finding underlying similarities between them using combinations of the following information:
\begin{itemize}[leftmargin=8em]
    \item[Text (T):] Each document contains $k_d$ word tokens from a vocabulary of $V$ word types ($M=\sum_d k_d$ is the total number of word tokens).
    \item[Hyperlinks (H):] Documents are linked to each other by building a (directed) graph or network.
    \item[Metadata (M):] Documents are classified by tags or other metadata.
\end{itemize}

These characteristics are typical for textual data and networks. Here we explore three types of such datasets, summarized in Tab.~\ref{tab:data-summary}. The main dataset we use to illustrate our results was extracted from the English Wikipedia, where the documents are articles (in scientific categories), the text is the content of the articles, hyperlinks are links between articles contained in the text, and metadata are tags introduced by users to classify the articles (categories). In our main example, we selected hundreds of articles in one of three scientific categories of Wikipedia (see Appendix~1 for details).  %\ref{section:data_collection}
 Our main findings are confirmed in a second Wikipedia dataset (obtained choosing different scientific categories), in a citation dataset (documents are scientific papers, hyperlinks are citations, the text is extracted from the title and abstract, and metadata are scientific categories), and in an E-mail dataset (documents are all E-mails from the same user, hyperlinks correspond to E-mails sent between users, and the text is the content of the E-mails). These results and further details of the data are presented in the Supplementary Information (see SI-Sec.~1).

\begin{table}[h!]
\caption{Summary of the datasets used in this paper.}
\begin{tabular}{|l|cccc|}
  \toprule
  Nodes & Wikipedia Dataset in &  Wikipedia Dataset in&E-mail Dataset in& Citation dataset in\\
~  & Manuscript & SI& SI& SI\\
\hline
    Documents & 120 & 316 & 4,894 & 2,542\\
    Word Types & 11,545 &  16,344& 66,088 & 7,677  \\
    Metadata Tags   & Physics, Maths, & Statistics, Maths,  & 0 &52 Categories\\
    ~ &  Biology & Electrical Engineering && \\
\bottomrule
\toprule
    Edges &  &&& \\
\hline    
    Hyperlinks  & 309 & 1,530 & 18,005 & 4,590    \\
    Word Tokens & 155,093 & 321,147 & 761,179 & 116,889 \\
    Tag Labels  & 120 & 316 & 0 & 2,542 \\
\bottomrule
\end{tabular}%
\label{tab:data-summary}
\end{table}

\subsection{Data as Networks}\label{sec.data}

The data described above can be represented as multilayer networks. 
The Hyperlink layer is the most obvious one, where documents are nodes and the hyperlinks are directed edges. The Metadata layer is built by a bipartite network consisting of metadata tags and documents as nodes, whereby undirected edges correspond to documents containing a given metadata-tag. Finally, the Text layer is obtained by restricting the text analysis to the level of word frequencies (bag-of-words) and then considering the bipartite network of word (types) and documents, where the edges correspond to word tokens (i.e., the count of how often a word type appears in a document). While word-nodes and metadata tags appear only in the text and the metadata layer, all layers have document nodes in common. The novelty of our multilayer approach, in comparison to other approaches using multilayer networks, is the inclusion of the text layer. The importance of using a bipartite multigraph layer~\cite{larremore} to represent the text, instead of alternative ``word networks''~\cite{Lancichinetti2015,Amancio2,Leydesdorff2017}, is that it contains the complete information of word occurrence in documents and allows for a formal connection to topic-modelling methods~\cite{Gerlach2018,Karrer2011}.

\begin{figure}[h!]
  \centering
  \includegraphics[width=8cm]{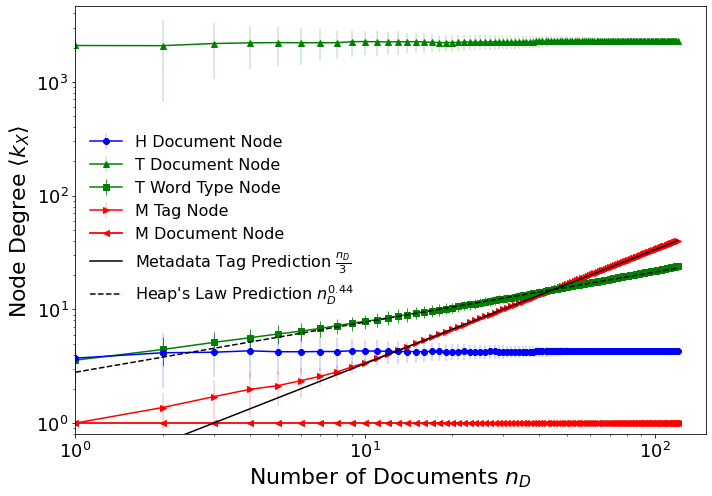} 
  \caption{
%  \textbf{Scaling growth of each class of node.} 
    \textbf{Scaling of average degree for each class of node reveals sparse and dense layers.} 
  Scaling of the average degree $\langle k_X \rangle$ with the number of documents $n_D$ depends on the node types $X$. The average degree was computed over all nodes of the same type (see legend, where H,T,M indicates the layer) in a sample of $n_D$ documents from  dataset. The symbols (error bars) are the average (standard deviation) over multiple random samples of documents. The prediction for the degree of word types using Eq.~(\ref{eq.scaling}) is also plotted for reference.
  }\label{fig:corpus-density}
\end{figure}

We now investigate the properties of the multilayer network described above, based on known results in networks and textual data. 
The most striking feature of this network is that the size of the different layers varies dramatically and scales differently with system size. A first indication of this lack of balance is seen by looking at the number of edges shown in Tab.~\ref{tab:data-summary}: the number of edges in the text layer (i.e. word tokens) is substantially larger than the number of nodes or edges in all of the other layers. Such an imbalance is expected in all datasets in which the same type of data as outlined in Sec.~\ref{sec.setting} is present. To see this we investigate in Fig.~\ref{fig:corpus-density} how the average degree $\langle k_X \rangle$ (number of edges/ total number of nodes) of the different node types~$X$ scale with the number of documents $n_D$ (which plays the role of system size). For the document nodes in the Hyperlink layer and the Text layer we see a constant average degree, typical of sparse networks. The Metadata layer yields a trivial scaling linear with $n_D$ as in dense networks because each document has one edge to a metadata node. More interestingly, the average degree of the word type nodes in the Text layer, $\langle k_V \rangle$, shows a growth that scales as
\begin{equation}\label{eq.scaling}
    \langle k_V \rangle \sim n_D^\gamma,
\end{equation}
with $0 < \gamma <1$. This is between the usual limits expected for sparse ($\gamma=0$) and dense ($\gamma=1$) networks. 

We now explain the observation in Eq.~(\ref{eq.scaling}) in terms of properties of text in general. More specifically, the type-token relationship in texts follows Heaps' law~\cite{Herdan1960, Heaps1978,Altmann2015}, which states that the number of word types $V$ scales with the word tokens $M$ as
\begin{equation}\label{eq.heaps}
V \sim M^{\beta},
\end{equation}
whereby $0<\beta<1$ is the parameter of interest. The average degree is obtained as $\langle k_V \rangle = M/V$ and $n_D \propto M$ (where the proportionality constant is the average size of Wikipedia articles, in word tokens). Combining this with Eqs.~(\ref{eq.scaling}) and (\ref{eq.heaps}) we obtain that $\gamma = 1-\beta$. From the data used here, we estimate a Heaps' exponent $\beta = 0.56$, that leads to a prediction of $\gamma=0.44$. This prediction is shown as a dashed line in Fig.~\ref{fig:corpus-density} and is in good agreement (for large $n_D$) with the average degree  of word nodes.

\subsection{Stochastic Block Models}\label{sec.sbm}

To achieve our goal of clustering documents and identifying topics considering multiple type of datasets simultaneously, we need to explore statistical patterns in the connectivity of the multilayer networks discussed above. This can be obtained using Stochastic Block Models (SBMs). The choice of SBM is based on the existence of a successful computational and theoretical framework, reviewed in Ref.~\cite{Peixoto2019bayesian}, that can be applied to networks with the characteristics needed in our problem:  different types of networks (directed, bipartite, and multi edges), multilayer networks~\cite{Peixoto2015Mesoscale}, and accounting for key ingredients to detect communities (e.g.,  degree correction and a nested/hierarchical generalizations \cite{Peixoto2014Hierarchical}). Our previous analysis of bipartite word-document networks using this framework has outperformed traditional topic modelling approaches~\cite{Gerlach2018}.

SBMs are a family of random-graph models that generate networks with
adjacency matrix $A_{ij}$ with probability $P(\bm{A}|\bm{b})$, where the
vector $\bm{b}$ with entries $ b_i \in \{1, \cdot \cdot \cdot, B\}, $
specifies the membership of nodes $i=1,\cdot \cdot \cdot, D$ into one of
$B$ possible groups. For our multilayer network design -- developed for
the three types of data (H,T,M) as discussed in Sec.~\ref{sec.data} --
we fit the SBM framework to each layer combining them by constraining
document groups to be the same across all layers, i.e. with a joint
probability
\begin{equation}
  P(\bm{A}_{\text{H}}, \bm{A}_{\text{T}}, \bm{A}_{\text{M}}|\bm{b}) = P(\bm{A}_{\text{H}}|\bm{b})P(\bm{A}_{\text{T}}|\bm{b})P(\bm{A}_{\text{M}}|\bm{b}),
\end{equation}
where $\bm{A}_{\text{H}}$, $\bm{A}_{\text{T}}$ and $\bm{A}_{\text{M}}$
are the adjacency matrices of each respective layer. In each individual
layer, edges between nodes $i$ and $j$ are sampled from a Poisson
distribution with average~\cite{Karrer2011}
\begin{equation}
  \theta_i\theta_j\omega_{b_i,b_j}
\end{equation}
whereby $\omega_{rs}$ is the expected number of edges between group r
and s, $b_i$ is the group membership of node $i$, and $\theta_i$ is
overall propensity with which a node is selected within its own
group. Non-informative priors are used for the parameters $\bm{\theta}$
and $\bm{\omega}$ and the marginal likelihood of the
SBM is computed as~\cite{Peixoto2017Nonparametric}
\begin{equation}
P(\bm{A}|\bm{b}) = \int P(\bm{A}|\bm{\omega}, \bm{\theta}, \bm{b})P(\bm{\omega},\bm{\theta}|\bm{b})d\bm{\theta}d\bm{\omega},
\end{equation}
Based on this, we consider the overall posterior distribution for a
single partition conditioned the edges on all layers~\cite{Hric2014}
\begin{equation}\label{eq.posterior}
  P(\bm{b}|\bm{A}_{\text{H}}, \bm{A}_{\text{T}}, \bm{A}_{\text{M}}) =
  \frac{P(\bm{A}_{\text{H}}|\bm{b})P(\bm{A}_{\text{T}}|\bm{b})P(\bm{A}_{\text{M}}|\bm{b})P(\bm{b})}{P(\bm{A}_{\text{H}},
  \bm{A}_{\text{T}}, \bm{A}_{\text{M}})}.
\end{equation}
With this approach, not only the words but also the documents are now clustered into categories.
We implement the inference using the package graph-tool~\cite{Graphtool, Peixoto2014MonteCarlo, Peixoto2020MergeSplit, Newman1999} (see SI-Sec. 2 for details and Ref.~\cite{Graphtool} for our codes).

\section{Application to Wikipedia data}\label{sec.3}

In this section we apply the methodology and ideas discussed above to the Wikipedia dataset  which contains articles classified by users in the categories Mathematics, Physics, and Biology. 
We are interested in comparing the outcomes and performance of the models discussed above applied to the different types of information in the data. We fit multiple variants of the multilayer SBM, whereby we choose different layers to be included in the model.

\subsection{Description Length}

The performance of each model can be measured by the extent to which a model succeeds in describing (compressing) the data. This can be quantified computing its description length (DL) ~\cite{Rissanen1978, Grunwald2007}
\begin{equation}\label{eq.dl}
  DL=-\log P(\bm{A}_{\text{H}}, \bm{A}_{\text{T}}, \bm{A}_{\text{M}},\bm{b}),
\end{equation}
which describes the information necessary to describe both the data and
the model parameters. From Eq.~\ref{eq.posterior}, we see that
minimizing the description length is equivalent to maximizing the
posterior probability $P(\bm{b}|\bm{A}_{\text{H}}, \bm{A}_{\text{T}},
\bm{A}_{\text{M}})$.

In Tab.~\ref{tab:DL-summary} we summarise the DL obtained for each model
in our dataset. It is quite clear that the DL of the models containing
the Text layer are much larger than those containing only the Hyperlink
and Metadata layers. This is a direct consequence of the large number of
word types in the data, when compared to documents or hyperlinks, the
lack of balance between the layers mentioned in
Sec.~\ref{sec.data}. This lack of balance between layers thus has
important consequences for the inference of partitions and our ability
to compare the different models. For instance, the effectiveness of the
multilayer approach could be evaluated by comparing the DL of the
multilayer model (e.g., DL of model $H+T$) to the sum of the DL of the
single-layer models (e.g., DL of model $H$ + DL of model $T$). In our
case this comparison is not very informative because the DL of the
combined model is dominated by the largest layer and the DL of the small
layer often lies within the fluctuations obtained from multiple MCMC
runs (see SI-Sec. 2). This reasoning suggests that the clustering of nodes would be dominated by the Text layer or, if the Text layer is excluded, by the Hyperlink layer which will dominate over the Metadata layer\footnote{In the Metadata layer, we found that the metadata tags will form a trivial single group as there is insufficient evidence for the model to construct more than one group. Therefore, we constrained metadata tags to be in separate groups to ensure that they provide additional information to the models being fitted.}. However, we will see below that there are still significant and meaningful differences in the clustering of nodes obtained using different combinations of layers. This happens because the inference problem remains non-trivial because the DL landscape contains many distinct states with similar values in the DL so that even small effects due to the $H$ and $M$ layers can affect the outcome.

\begin{table}
\caption{Description length for each combination of layers in the multilayer stochastic block model. We compute the average description length (DL), Eq.~(\ref{eq.dl}), for each model class alongside the standard deviation over multiple MCMC runs. We also retrieved the minimum DL (MDL) over all the runs. The DL of the Text layer exceeds the Hyperlink and Metadata layer by several orders of magnitude, thus contributing the most to the Hyperlink + Text model.
}
\begin{tabular}{ |c|c|c|c|} 
\toprule
Model & Layers & DL & MDL \\
\hline
H & Hyperlink & 1,135 (0)  & 1,135\\
T & Text & 257,775 (471) &  256,973\\
M & Metadata & 76 (0) & 76\\
H+M & Hyperlink+Metadata & 1,295 (18)   & 1281\\
H+T & Hyperlink+Text & 270,230 (2,228)  & 267,102\\
H+T+M & Hyperlink+Text+Metadata & 282,560 (624)  & 281,133\\
\bottomrule
\end{tabular}
\label{tab:DL-summary}
\end{table}

\subsection{Qualitative comparison of groups of documents}

\begin{figure}[h!]
\begin{subfigure}[ht!]{3.1cm}
\centering
\includegraphics[height=3cm]{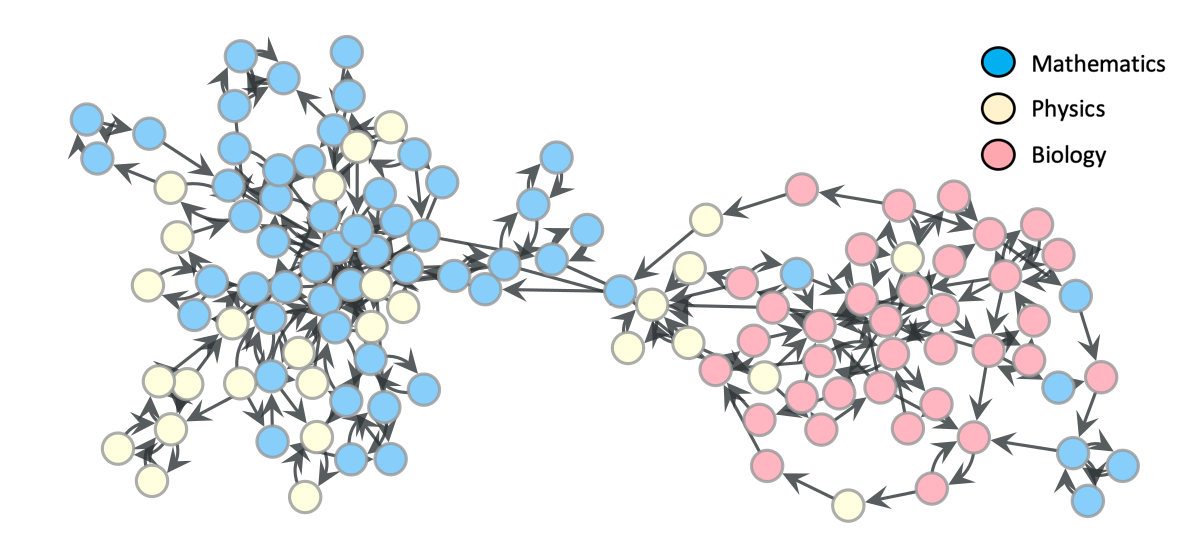}
\caption{Wikipedia Labels}%
\end{subfigure}
\hfil
\hfil
\hfil
\hfil
\hfil
\begin{subfigure}[ht!]{5cm}
\centering
\includegraphics[height=3cm]{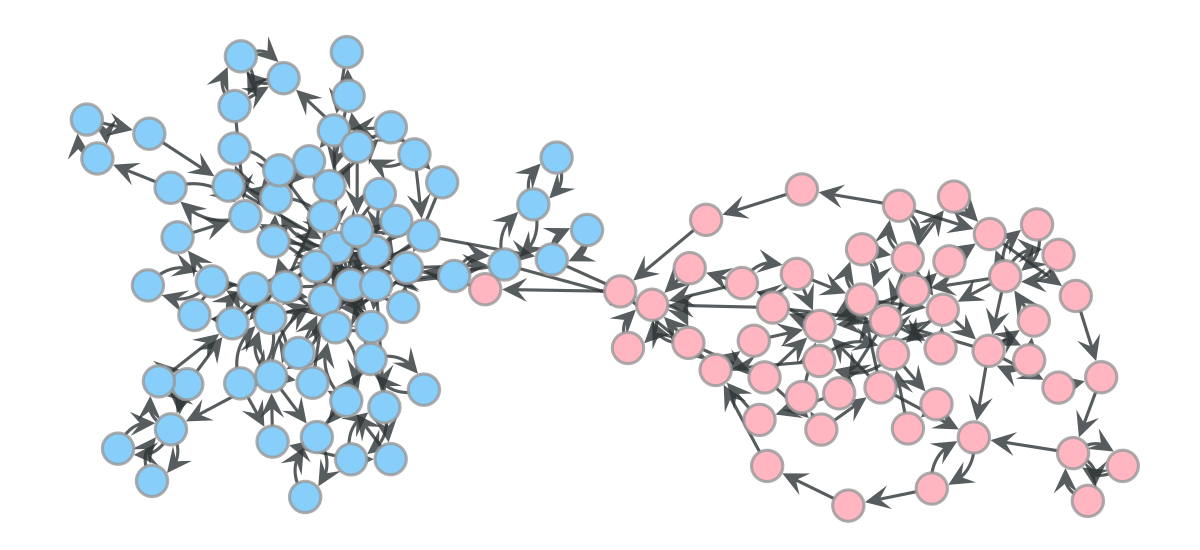}
\caption{Hyperlink Model, $\operatorname{\sigma} = 0.0$}%
\end{subfigure}
\newline
\begin{subfigure}[ht!]{4.3cm}
\includegraphics[height=3cm]{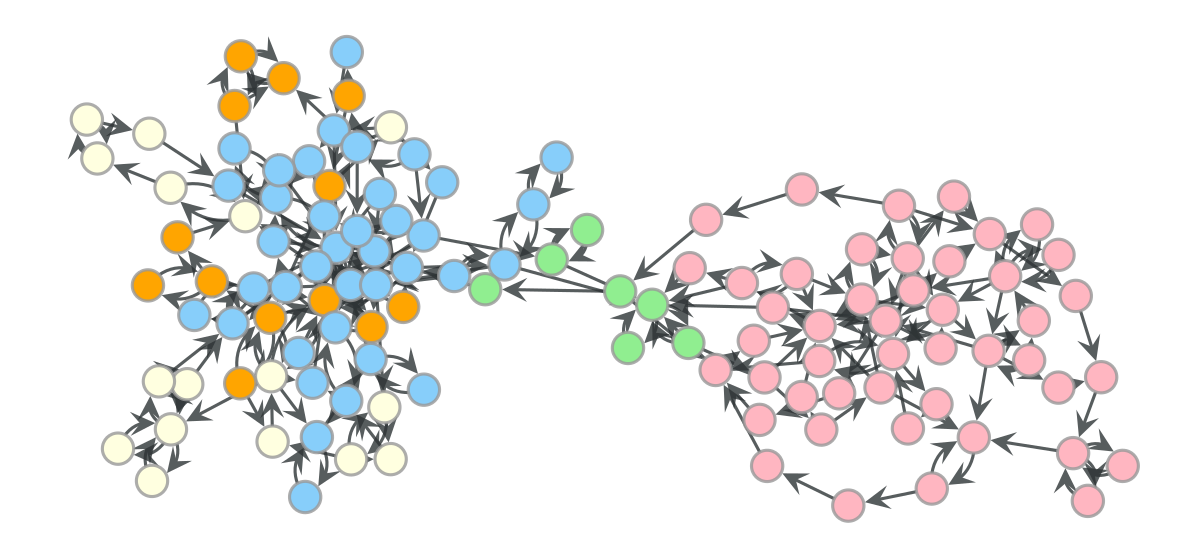}
\caption{Text Model, $\operatorname{\sigma} = 0.27$}%
\end{subfigure}
\hfil
\hfil
\hfil
\hfil
\hfil
\begin{subfigure}[ht!]{6cm}
\includegraphics[height=3cm]{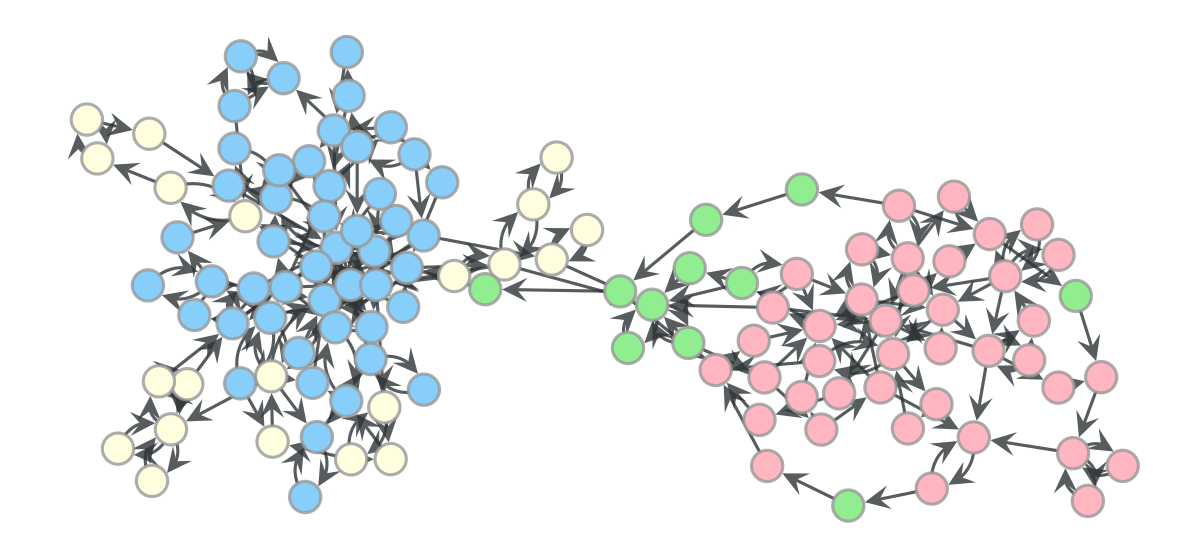}
\caption{Hyperlink + Text Model, $\operatorname{\sigma} = 0.088$}%
  \end{subfigure}
\caption{\textbf{ Different models lead to different partitions of Wikipedia articles into communities.} The network corresponds to Wikipedia articles (nodes) and hyperlinks (edges). The colour of the nodes corresponds to groups of document nodes: (a) from the Wikipedia labels (with annotations Mathematics, Physics, and Biology); (b-d)  communities found from our model using different datasets (layers) and the consensus partition (see App.~2). The uncertainty $\sigma$ of the reported partitions is quantified by $0 \le \sigma \le 1$, defined in Eq.~(\ref{eq.sigma}) below.
}

\label{fig:partition-consensus}%
\end{figure}
Community detection methods aim to find the partition of the nodes that best captures the structure in the network in a meaningful way whilst being robust to noise \cite{Peixoto2019bayesian,Hastings2006}.  We thus evaluate the different models by comparing the resulting partitioning of documents \cite{Hric2014}.
Specifically, we fit the Hyperlink, Text, and Hyperlink + Text model and obtain a best partition from combining multiple samples from the posterior $P(\bm{b}|\bm{A})$ for each model to construct a point estimate, which utilises the different parts of the posterior distribution. We then project the group membership onto the Hyperlink layer (which only contains document-nodes) and retrieve the consensus partition alongside the uncertainty of the partition \cite{Peixoto2020} (see Appendix~2 for details).

Our results are shown in Fig.~\ref{fig:partition-consensus} and reveal that our model is successful in retrieving different meaningful groupings of the articles depending on the available data (i.e. layers included in the model). We first notice that the classification of articles made by users -- panel (a), Wikipedia label -- group articles in Mathematics and Biology that are strongly linked with each other (through hyperlinks), whereas Physics articles appear intertwined in between them. When we infer the partition of nodes based only on the hyperlink network -- panel (b), Hyperlink model -- we obtain that our model obtains 2 groups and it is quite confident about it (uncertainty is zero, $\sigma = 0.$). This partition resembles the partition based on Wikipedia labels. When the documents are partitioned based on their text -- panel (c), Text Model--, a richer picture emerges. There is a large community that resembles closely the documents classified as Biology and one of the communities obtained using the hyperlinks layer. However, the remaining documents (most of the Mathematics and Physics articles) are now grouped in $4$ categories (i.e., $5$ communities in total) which are still linked to each other but more loosely than before (even though Fig.~\ref{fig:partition-consensus} shows the Hyperlink network, the Hyperlinks were not used to group documents in panels a) and c)). Finally, when hyperlinks and text are used simultaneously -- panel (d) -- $4$ communities are found, which resemble the previous ones but that also show important distinctions. This demonstrates that even if the Text layer dominates the description length, there are noticeable differences in the inferred partitions when using the hyperlinks in addition to text for clustering documents.    

We now argue that the more nuanced classification of documents obtained with the Text and Hyperlink + Text models are qualitatively meaningful. For example, we can see a cluster of 5 (Physics) nodes in the bottom left of the Hyperlink model that was not identified as a separate group, but it is now picked up in the Text and Hyperlink + Text model. This cluster of nodes include Wikipedia articles on the Josephson effect, macroscopic quantum phenomena, magnetic flux quantum, macroscopic quantum self trapping, and quantum tunnelling. Even more strikingly, in the bottom of the network there is a lone (Physics) green node surrounded by (Biology) red nodes which corresponds to the Wikipedia article on isotopic labelling (a technique in the intersection of Physics and Biology). In traditional community detection methods, which use link information as an indicator of groups, such a node would be in the community of its surrounding neighbours. However, in the Hyperlink + Text model, we are able to detect the uniqueness of such a node.

\subsection{Quantitative comparison between different models}

In the example discussed above it was clear that the different models yielded different yet related partitions of Wikipedia articles. In order to quantify the similarity of the results of the different models, we performed a systematic comparison of the partitions generated by multiple runs of each model and computed their similarities using the maximum overlap partition (Fig.~\ref{fig:corpus-2-partition-overlap-matrix}, see Appendix~2 for details). The results show   that the partitions generated by the Hyperlink + Text model is most similar to the Text model. Similar results are obtained in our alternative datasets -- see SI-Sec.~1 -- and using the normalised mutual information (NMI) as an alternative dissimilarity measure --see SI-Sec.~3. 

\begin{figure}[b!]
\includegraphics[width=10cm]{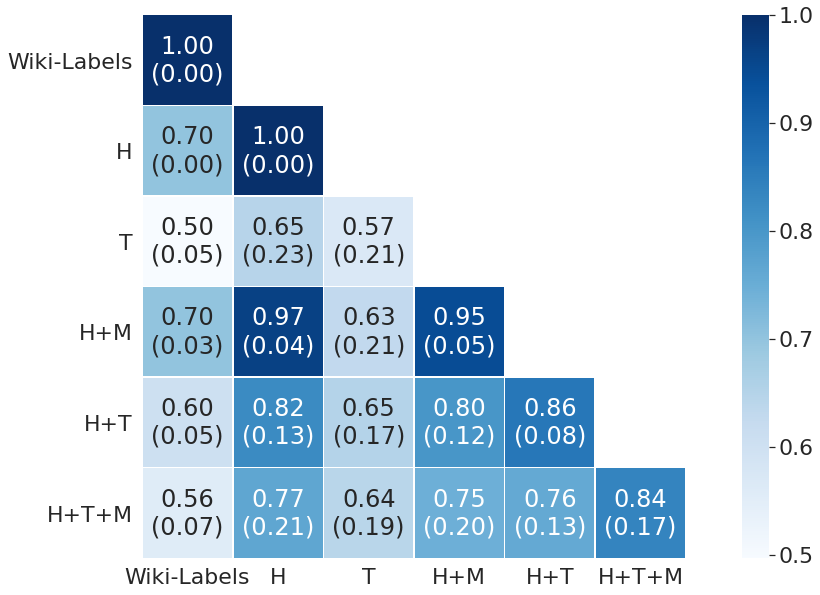}
\caption{
\textbf{Maximum partition overlap of the consensus partitions between the model classes.} The average and standard deviations of the maximum partition overlap between and within different models.
}\label{fig:corpus-2-partition-overlap-matrix}
\end{figure}

We also compare the Hyperlink and Hyperlink + Text model in terms of their ability to predict missing edges \cite{Guimera2009,Valles-Catala2018} (see Appendix~3 for details on our method). %\ref{ssec.linkprediction}
We found that the Hyperlink+Text model has an Area-Under-Curve (AUC) score of $ 0.63 \pm 0.06$ (average $\pm$ standard deviation) and the Hyperlink model has $0.54 \pm 0.02$, with the difference being statistically significant ($p=0.0013$, using a 2-sample t-test). This confirms that the multilayer approach proposed here is successful in retrieving existing relationships that are missed in the network-only approach. 

\subsection{Lack of balance in the Hyperlink-Text model}

\begin{figure}[h!]
  \centering
  \includegraphics[width=10cm]{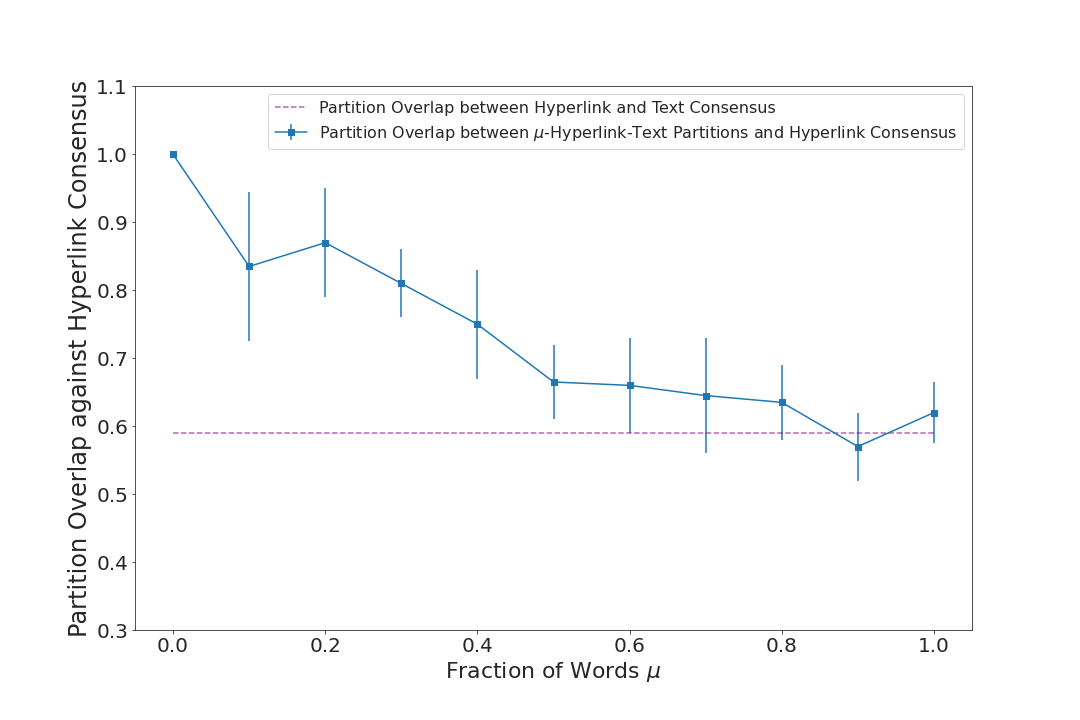} 
  \caption{\textbf{Text layer determines the partitions obtained in the multilayer model (Hyperlink + Text).} 
  Similarity (overlap of consensus partition) between Hyperlink partition and $\mu$-Hyperlink-Text partition as a function of the subsampling parameter $\mu$ where $\mu=0$ ($\mu=1$) corresponds to the case with all (none) of the word tokens removed in the Hyperlink-Text model.
  For a given value of $\mu$, a random fraction $1-\mu$ of the words were removed and the Hyperlink + Text model was then fitted for multiple iterations. The consensus partition was then computed for the Hyperlink + Text model and its partition overlap with Hyperlink model.  
  A higher sub-sampling of text (i.e. smaller values of $\mu$) results in the consensus partition between the Hyperlink + Text and Hyperlink model having a high degree of overlap.}
\label{fig:partition-overlap}  
\end{figure}

The results of the previous sections are strongly influenced by the lack of balance in Hyperlink + Text model, as discussed in Sec.~\ref{sec.2}. To further illustrate this point, here we artificially reduce the unbalance of the multilayer network by sampling a fraction $\mu$ of word tokens before fitting a Hyperlink + Text model. We expect that, as we increase the fraction of words $\mu$, the Text layer will increasingly dominate the inference. This expectation is confirmed in Fig.~\ref{fig:partition-overlap}, which shows that for $\mu \geq 0.6$, the partition overlap of the $\mu-$hyperlink-text model is statistically indistinguishable from the partition overlap obtained using the Text-only model. That is, we see that  the Text layer  dominates the inference in the $\mu-$Hyperlink + Text for $\mu \geq 0.6$. However, as discussed above, the effect of the hyperlink layer can lead to different consensus partition. 

\subsection{Topic Modelling: Groups of Words}
Since our approach provides a clustering of all nodes, we not only group documents but also words. The groups of word (types) can be interpreted as the topics of the documents linked to them, showing that our framework simultaneously solves the traditional problem of topic modeling~\cite{Blei2003,Lancichinetti2015}. Below we show the topics obtained in our Wikipedia dataset, as an example of our generic topic-modelling methodology.

In the consensus partition of the Hyperlink-Text network (see Fig.~\ref{fig:partition-consensus}) we found 12 topics (groups of word types). The most frequent words in each of these topics is shown in Tab.~\ref{tab:HT-topics}. Qualitatively, we see that topics are often composed of semantically related words, e.g. topics 1 and 3 contain a large number of key words associated to Biology whilst topics 5 and 10 contains a large number of jargon related to Physics.

\begin{figure}[h!]
  \centering
  \includegraphics[width=10cm]{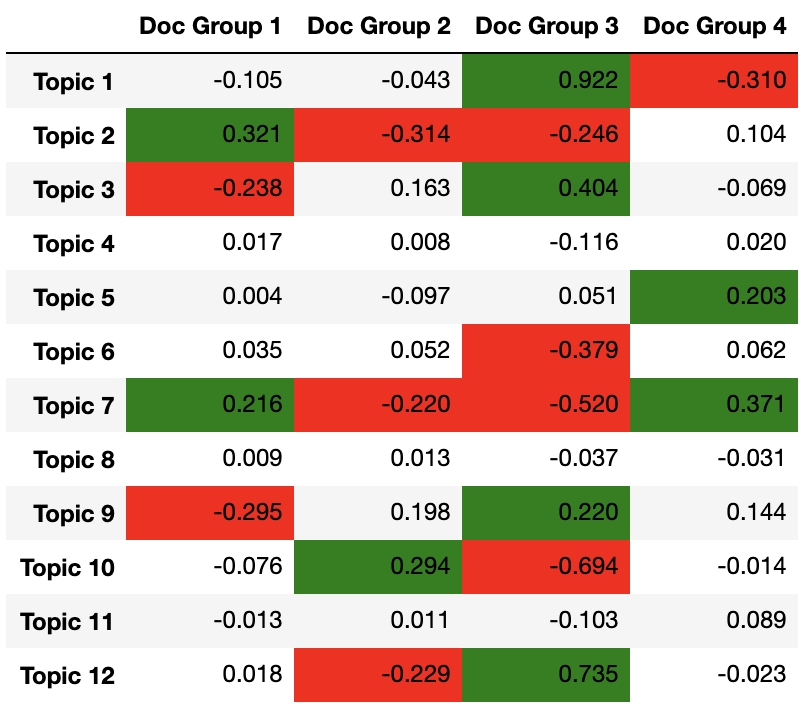} 
  \caption{\textbf{Normalised contribution of topics to the group of documents.} The normalized measure~(\ref{eq:normalised_topic}) was computed for all 4 document groups and 12 word groups (topics).
  We set a threshold of $\tau_i^t \geq 0.2$ ($\tau_i^t \leq -0.2$) to define if a topic is over- (under-) represented in a document group $d$. 
  }
\label{Fig:Group-Topic-Representation}  
\end{figure}

We now discuss the topical composition of (groups of) documents. Let $T=B_V$ be the number of topics and $B_D$ be the number of document groups, then the mixture proportion of topic $t =1,\ldots,T$ in document group $i =1,\ldots,B_D$ is given by 
\begin{equation}\label{eq.fi}
    f_{i}^{t} = \frac{n_{i}^{t}}{\sum_{t'=1}^{T}n_{i}^{t'}},
\end{equation}
where $n_{i}^{t}$ is the number of word tokens in topic $t$ that appeared in documents $d$ in document-group $i$. The results obtained for the four document groups are shown at the bottom of Tab.~\ref{tab:HT-topics}. Interestingly, topic $4$ cannot be identified with any specific group of documents. This suggests that the words in this topic are similar to so-called stopwords, a pre-defined set of common words considered uninformative which are typically removed from the corpus before any model is to be fitted in order to improve the model \cite{Haddi2013}. This is consistent with the finding of Ref.~\cite{Gerlach2018} that SBMs applied to word-document networks were able to automatically filter stop words by grouping them into a ``topic'' that is well connected to all documents. Our findings suggest that the same is true for multilayer models and that our approach is robust against the presence of stopwords. In fact, this stopword topic is responsible for a large fraction ($40\%$) of the topic-proportion for all groups of documents. The underlying reason for this is the higher frequency of these words, which (due to Zipf's law) dominate the weights of the topic mixture models ~\cite{Altmann2017}. To overcome this feature, and assess the over- or under-representation of topics more rigorously,  we account for the overall frequency of occurrence of words in topics $t$ as
\begin{equation}
\langle f^{t} \rangle = \frac{\sum_{i=1}^{B_D}n_{i}^{t}}{\sum_{t=1}^{T}\sum_{j=1}^{B_D} n_{j}^{t}},
\end{equation}
and define the normalised value of the mixture proportion of topic $t$ in document group $i$ as
\begin{equation}
    \tau_{i}^{t} = \frac{f_{i}^{t} - \langle f^{t} \rangle}{\langle f^{t} \rangle}.
    \label{eq:normalised_topic}
\end{equation}
This normalised measure has an intuitive interpretation: $\tau_{i}^{t} > 0$ ($\tau_{i}^{t} < 0$) implies that topic $t$ is over-represented (under-represented) in document group $d$.
In Fig.~\ref{Fig:Group-Topic-Representation}, we show $ \tau_{i}^{t} $ for the 12 topics and the 4 document groups, providing a much clearer view on the connection between topics and groups of documents. For example, we see that document group 2 (articles labelled as Physics) has a large over-representation of topic 10, which corresponds to the Physics topic whilst being underrepresented in document group 2 (articles labelled as Biology). 
Looking at the model's newly proposed document group (group 4) we see that it has an over-representation from topics 7 and 5 (and in a less extent from topics 2, 9, and 11), confirming its hybrid category.
  
\begin{table}[h!]
\caption{Groups of word types as topics. Upper table: the $20$ most frequent words in the 12 topics (word groups) found in the consensus partition of the Hyperlink + Text model. Lower table: the topic proportion~(\ref{eq.fi}) of the four groups of documents.}
  \resizebox{1 \textwidth}{0.15 \textheight}{%
\begin{tabular}{ |c|c|c|c|c|c|c|c|c|c|c|c|c|}
\hline
Group & Topic 1 & Topic 2 & Topic 3 & Topic 4 & Topic 5 & Topic 6 & Topic 7 & Topic 8 & Topic 9 & Topic 10 & Topic 11 & Topic 12\\
\hline
& occur &   chemical &  protein & one & system & group & transform & action & happen & quantum & every & factor\\
& specific & loop &     cell &  also &  time &  space & law  &  translation & frequency & equation & basis & expression\\
& within & electron & site & use & state & define & critical & class & fast & transformation & unique & region\\
& process & transition & activity & form & work & field & magnetic & line & output & theorem & degree & molecular\\
& produce & chemistry & pathway & function &    second & point & matter & measure & input & symmetry & positive & sequence\\
& increase & atom & cellular & give & effect & theory & scale & central & algorithm & dimension & open & information\\
& mechanism & stable & organism & two & phase & value & plane & primary & digital & denote & close & end\\
& cause & ion & amino & may & interaction & product & spectrum & unknown & & classical & fix & molecule\\
& acid & pattern & enzyme  & however & potential & constant & volume & interior && let & reference & domain\\
& target & crystal & synthesis & example & could & vector & temperature & surface &&    mechanic & exact & signal\\
& step & reach & species & result & derive &    physic &    axis &  side &  &       coordinate & closed &   alternative\\
& bind& prediction & tissue & call&difference& physical& statistical& center& & lie& restrict & bond\\
& gene& label&proteins& first& full& energy& sign& flow& & real& picture& family\\
& control& compound& cancer& different&     free& element& effective& copy& &       mathematics&    infinitesimal&  interact\\
& human& reaction& activate& know& hold& particle& symbol& block& &         geometry& index& cycle\\
& encode& crystallography&  genetic& make& observe& parameter& speed& configuration & &     operator& interpretation&   biological\\
& body& oxygen& membrane& number& research& whose& electric& read& & differential& formulate&   life\\
& dna& solid& release& structure& propose& mathematical& equilibrium& alternate& &      representation& equivalently& translate\\
& rate& unstable& mutation& show& year& linear& unchanged&  table& & invariant&     maximal &   structural\\
& formation&    atomic& regulation& include& paper& matrix& assumption& transport& &        generalize& scheme& biology\\
\hline
\hline
1 (Mathematics) & 0.1481 & 0.0213 & 0.0444 & 0.400 & 0.0578 & 0.0722 & 0.0510 & 0.0127 & 0.00163 & 0.147 & 0.0105 & 0.0338\\
2 (Physics) & 0.138 & 0.0409 & 0.0291 & 0.404 & 0.0643 & 0.0710 & 0.07940 & 0.0127 & 0.0009583 & 0.105 & 0.0102 & 0.0445\\
3 (Biology) & 0.297 & 0.0234 & 0.0536 & 0.351 & 0.0673 & 0.0426 & 0.0313 & 0.0121 & 0.00166 & 0.0347 & 0.00929 & 0.0759\\
4 (New Group)& 0.107 & 0.0342 & 0.0355 & 0.405 & 0.0771 & 0.0728 & 0.0895 & 0.0122 & 0.00156 & 0.112 & 0.0113 & 0.0428 \\ 
\hline
\hline
  \end{tabular}}
\label{tab:HT-topics}
\end{table}

\section{Discussion and Conclusions}

In this paper, we introduced and explored a formal methodology that combines multiple data types (e.g., text, metadata, links) to perform the common tasks of clustering and inferring latent relationships between documents in text analysis. The main theoretical advantage of our methodology is that it incorporates all the different types of data into a single, consistent, statistical model.  Our approach is based on an extension of multilayer Stochastic Block Models, that have been used previously to find communities in (sparse) complex networks and that is used here to perform text analysis (see Refs.~\cite{Zhu2013,Bouveyron2016} for alternative uses of SBMs for topic modelling). On the one hand, our method extends community-detection methods to the analysis of text in the presence of multiple data types, our main finding being that: (i) universal statistical properties of texts lead to different link densities at the different layers of the network; and (ii) that the word layer plays a dominant role in the inference of partitions. On the other hand, our method can be viewed as a generalized topic modelling method that incorporates meta-data and hyperlinks, labels the communities of documents by examining the proportion of topics, and builds on the previous connections between SBMs and Latent Dirichlet Allocation~\cite{Gerlach2018, Karrer2011}.

Our investigations on four different datasets show consistent results that reveal the potential and limitations of our approach. Our most important finding is that our methodology succeeds in using the multiple data types (e.g., a text layer) leading to more nuanced communities of documents and in increasing the ability to predict missing links. On the practical side, the lack of balance between the different layers poses challenges on how to evaluate the contributions of different layers because the description length obtained in the inference process is dominated by the text layer and variations obtained within the (Monte Carlo) inference process become larger than the contribution of alternative layers. This suggests further investigations on the role of unbalanced layers in multilayer networks, and how to deal with them within the proposed framework, as important steps to expand the success of complex-network methods to other classes of relevant datasets.

\newpage
\section*{Appendix}

\section*{Appendix 1: Wikipedia Data Collection and Preparation}
%\label{section:data_collection}
We used a snapshot of the Wikipedia data retrieved on the 5th of June, 2020. The following lists the data extraction and processing steps:

\begin{enumerate}
    \item {\it Data Retrieval}: We retrieved the Wikipedia articles and their content (metadata, text, link) through the MediaWiki API.\footnote{\url{https://en.wikipedia.org/w/api.php}} and parsing the Wikipedia dumps. \footnote{\url{https://dumps.wikimedia.org/}}
    \item {\it Network Formulation}: We constructed a network whereby each node represents a Wikipedia article and each (directed) edge represents hyperlinks between the Wikipedia articles. We removed any nodes with less than 2 outgoing links.
    \item {\it Retrieve Connected Component}: For ease of analysis, we extracted the largest connected component in the hyperlink network constructed.
    \item {\it Text Processing}: We process the Wikipedia text data through both tokenization and lemmatization using NLTK~\cite{nltk}.
    %Wordnet corpora  which is a lexical database of semantic relationships between words. 
\end{enumerate}

The resultant dataset is available in our repository~\cite{Graphtool}.

\section*{Appendix 2: Maximum Overlap and Consensus Partition} 

The maximum overlap between partitions measures the similarities between sets of partitions. The maximum overlap between partitions $\bm{x}$ and $\bm{y}$ is given by 
\begin{equation}
    w(\bm{x}, \bm{y}) = \argmax_{\bm{\mu}}\sum_{i}\delta_{x_{i}, \mu(y_{i})},
\end{equation}
where $\bm{\mu}$ is a bijective mapping between the group labels \cite{Peixoto2020}.

The normalized maximum overlap between partitions $\bm{x}$ and $\bm{y}$ is given by 
\begin{equation}
    w(\bm{x}, \bm{y}) = \frac{\sum_{i}\delta_{x_{i}, y_{i}}}{N},
\end{equation}
 where N is the number of nodes and lies in the unit interval [0,1].

Given multiple partitions, we also wish to extract a consensus partition $\hat{\bm{b}}$ which has the maximal sum of overlaps with all the partitions. Such a consensus partition can be obtained through the double maximization of the set of equations: 
\begin{equation}
    \hat{b}_{i} = \argmax_{r}\sum_{m}\delta_{\mu_{m}(b_{i}^{m}),r}
\end{equation}
\begin{equation}
    \bm{\mu}_{m} = \argmax_{\bm{\mu}}\sum_{r}m_{r,\mu(r)}^{(m)},
\end{equation}
where $\bm{\mu}$ is a bijective mapping between the group labels and $m_{r,\mu(r)}^{(m)}$ is the contingency table between $\hat{\bm{b}}$ and partition $\bm{b}^{(m)}$. An iterative procedure is then carried out on the set of equations until no further improvement is possible. The uncertainty $\sigma$ of the consensus partition  obtained from $M_p$ partitions can be quantified as~\cite{Peixoto2020,Graphtool}
\begin{equation}\label{eq.sigma}
\sigma = 1 - \frac{1}{N M_p} \sum_i \sum_m \delta_{\mu_{m}(b_{i}^{m}),\hat{b}_i}.
  \end{equation}

\section*{Appendix 3: Supervised Learning Via Link Prediction}%\label{ssec.linkprediction}

A supervised learning approach to select the best model can be done through the task of link prediction \cite{Valles-Catala2018, Clauset2008}. Let $\bm{A}^{O}$ be the observed network and $\delta \bm{A}$ be missing or spurious edges. The desired posterior distribution of missing entries $\delta \bm{A}$ conditioned on the observed network $\bm{A}^{O}$ can be computed as 
\begin{equation}
    P(\delta \bm{A}|\bm{A}^{O}) = \frac{\sum_{\bm{b}}P(\bm{A}^{O} \cup \delta \bm{A}|\bm{b})P(\bm{b}|\bm{A}^{O}) }{P(\bm{A}^{O}|\bm{b})}.
    \label{eq:posterior_link}
\end{equation}
However, as the normalization constant is difficult to obtain, the numerator of Eq.~\ref{eq:posterior_link} can be computed by sampling partitions from the posterior and then inserting or deleting edges from the graph and computing the new likelihood. As a result, we therefore may compute the relative probability between specific sets of alternative predictive hypotheses $\{\delta \bm{A}_i\}$ through the likelihood ratios ratio 
\begin{equation}
    \lambda_i = \frac{P(\delta \bm{A}_i|\bm{A}^{O})}{\sum_jP(\delta \bm{A}_j|\bm{A}^{O})}.
\end{equation}

We can compute the area under curve (AUC) of the  receiver operating characteristic curve to evaluate the SBM's classification abilities. Furthermore, given two sets of AUCs from two different models, we can compare the models' performance by computing the t-statistic for a null model with zero mean for the difference in AUC which is given by 
\begin{equation}
    t_{\Delta AUC} = \frac{\langle \Delta AUC \rangle}{\sigma_{\Delta AUC}/\sqrt{n}},
\end{equation}
where $\langle \Delta AUC \rangle$, $\sigma_{\Delta AUC}$, and n are the mean, standard deviation and size of the population.

%%%%%%%%%%%%%%%%%%%%%%%%%%%%%%%%%%%%%%%%%%%%%%
%%                                          %%
%% Backmatter begins here                   %%
%%                                          %%
%%%%%%%%%%%%%%%%%%%%%%%%%%%%%%%%%%%%%%%%%%%%%%

\begin{backmatter}

  \section*{Additional Material}
The supplementary Material is available at \url{https://epjdatascience.springeropen.com/articles/10.1140/epjds/s13688-021-00288-5#Sec16}. Codes and data are available at \url{https://topsbm.github.io}.

\section*{Acknowledgements}
    Funding from The University of Sydney was received through the CTDS incubator scheme.

\end{backmatter}
\end{document}